\documentclass[doublespacing]{elsart}
\usepackage{natbib,amsmath}

\journal{Theoretical Population Biology}

\begin{document}

\begin{frontmatter}

\title{Limited resources and evolutionary learning may help to understand
the mistimed reproduction in birds caused by climate change}

\author[label1]{Daniel Campos\corauthref{cor}},
\corauth[cor]{Corresponding author.}
\ead{Daniel.Campos@uab.es}
\author[label2]{Josep E. Llebot} and 
\author[label2]{Vicen\c{c} M\'{e}ndez}

\address[label1]{School of Mathematics, Department of Applied Mathematics. The
University of Manchester, Manchester M60 1QD, UK.}

\address[label2]{Grup de F\'{\i}sica Estad\'{\i}stica. Departament de F\'{\i}sica.
Universitat Aut\`{o}noma de Barcelona, 08193 Bellaterra (Barcelona) Spain}

\newpage

\begin{abstract}
We present an agent-based model inspired by the Evolutionary Minority Game
(EMG), albeit strongly adapted to the case of competition for limited resources
in ecology. The agents in this game become able, after some time, to predict
the \textit{a priori} best option as a result of an evolution-driven learning process. We show that a self-segregated social structure can
emerge from this process, i.e., extreme learning strategies are always
favoured while intermediate learning strategies tend to die out. This result may
contribute to understanding some levels of organization and cooperative
behaviour in ecological and social systems. We use the ideas and results reported here to discuss an issue of current
interest in ecology: the mistimings in egg laying observed for some species of bird as
a consequence of their slower rate of adaptation to climate change in comparison with that
shown by their prey. Our model supports the hypothesis that habitat-specific constraints could
explain why different populations are adapting differently to this
situation, in agreement with recent experiments.
\end{abstract}


\end{frontmatter}
\newpage

\section{Introduction}

Minority games (Challet and Zhang, 1998), and more recently Evolutionary
Minority Games (EMG) (Johnson et. al., 1999a; Johnson et. al., 2000; de Cara
et. al., 2000; Johnson et. al., 2003; Hod and Nakar, 2002; Hod, 2003;
Sysi-Aho et. al., 2003; Johnson et. al., 1999b; Lo et. al., 2000), have
received widespread attention in recent years as a useful model to describe
competition for highly limited resources in complex systems, especially in
economics. These games are essentially based on a minority rule (Challet and
Zhang, 1998) according to which $N$ agents compete repeatedly for some
resources by choosing between two options A or B. Each agent makes its
choice, and those agents belonging to the less (most) frequently chosen
option are considered the winners (losers), so they are rewarded (fined).
So, the idea behind this game is that the agents must always try to be in
the minority: few individuals choosing the same option as yourself means
less competitors, and so it should be easier to obtain the resource. The
decisions taken by the agents are chosen according to a pool of strategies
available, and these strategies are based on the $m$ previous outcomes in
the game, as that information is assumed to be accessible to all of the
agents. To give a simple example, a specific strategy in a minority game
with $m=2$ has the form 
\[
S=\{(A,A)\rightarrow A,\text{ }(A,B)\rightarrow B,\text{ }(B,A)\rightarrow A,%
\text{ }(B,B)\rightarrow A\}. 
\]
This means that if the two previous winning options in the game were (A,A),
an agent following strategy $S$ will choose option A the next time; if the
last winning options were (A,B), that agent will choose B, and so on. At the
beginning of the game several strategies are assigned to each agent, and the
agent tends to choose from among them the strategy that gave better results
in the past; however, many different versions of the minority game exist,
where the rules that determine the strategies chosen by the agents are
different. Here, we will skip the minor details on the mechanisms of the
minority game, since that is outside the scope of the current work; an
exhaustive compilation of works on minority games can be found in
http://www.unifr.ch/econophysics.

In the evolutionary version (EMG) of the game (Johnson et. al., 1999a), all
the agents are assigned the same strategies but they can \textit{i)} follow
that given strategy with probability $p_{k}$ or \textit{ii)} do exactly the
opposite with probability $1-p_{k}$, where $p_{k}$ is different for each
agent (the subindex $k$ denotes the $k$th agent). Those agents performing
the worst (losing many times) are forced to change their value of $p_{k}$;
so, in the EMG there is an implicit learning process based on \textit{trial
and error. }As a consequence, the system tends towards an optimal
distribution of $p_{k}$ values for which the number of winners is as close
to $N/2$ as possible (note that, by definition, in a minority game the
number of winners cannot be higher than $N/2$). As reported in (Johnson et.
al., 1999a), the most striking result arising from the EMG is the natural
emergence of segregated behaviour: those agents that behave in an extreme
way ($p_{k}\rightarrow 0$ and $p_{k}\rightarrow 1$) perform better than
those with intermediate behaviour, so that the individuals tend to segregate
into two groups: those who always follow the given strategy and those who
never follow the strategy. From the point of view of complex systems, it has
been claimed that this result may help to understand some levels of
organization such as crowding (Johnson et. al., 2000; Cont and Bouchaud,
2000) and cooperation (de Cara, 2000), which are common in many social and
biological systems. Specifically, within the context of the EMG some authors
have coined the term \textit{unintentional }or \textit{indirect cooperation}
to illustrate the behaviour observed (Quan et. al., 2003; Hod and Nakar,
2004)\textit{.} This concept refers to the fact that in the EMG many agents
tend to behave similarly (either $p_{k}\rightarrow 0$ or $p_{k}\rightarrow 1$%
), but not consciously, but rather because the global winning probability is
higher that way. This is different from other games (for example, the
well-known Prisoner's Dilemma) where cooperation is a conscious option given
to the agents (Nowak et. al., 2004; Nowak 2006).

\section{Minority Games in ecology}

In general, minority games are helpful to describe multi-agent systems where
each agent (individual) is able to analyze the history of the system (i.e.,
the success of the different strategies used before) in order to make its
next decision. For this reason they have been especially designed and used
to explain the complex dynamics of some financial markets (Johnson et. al.,
2003), albeit some authors have stated that similar ideas could also hold
within an ecological context; probably the best example being foraging
behaviour (Hod and Nakar, 2002; Hod, 2003). However, as far as we know very
few real efforts have been made to extend minority games to ecological
situations. In (Tella et. al., 2001) the authors presented a model, inspired
by the rules of the minority game, to explore the \textit{colonial versus
solitary} behaviour in birds as a function of predation pressure, and some
discussions on the connection between minority games and ecological
evolution were provided in (de Cara et. al., 2000; Aho et. al., 2003).

The apparent lack of interest by ecologists in these games is probably due
to the fact that the most interesting and dramatic situations concerning
decision-making in animals are not well described by such concepts as 
\textit{trial and error }and \textit{pool of strategies} involved in
minority games. Instead, in ecology most of the interest lies in
understanding those situations where individuals perform just one or a few
critical decisions throughout their whole life (concerning, for example,
timing in reproduction or choice of habitat); these decisions have been
called 'fitness-critical actions' in a very recent work by Heesch and Little
(2006). Intuitively, decision-making in these 'fitness-critical actions'
follows quite simple mechanisms (compared to the complex rules of minority
games): the individuals need to use their skills or their experience to
predict the \textit{a priori }best option. By \textit{a priori }best option
we mean that option which would be the winning one in the case where half of
the agents choose A and the other half choose B. In the basic minority game
described above we have considered that the agents choose between two
identical options A or B, so there is no \textit{a priori }best option.
However, it is easy (and more realistic) to consider a game where A and B
are intrinsically different. For example, in the case of habitat selection,
individuals usually need to choose between different options with different
habitat qualities. Some individuals may be able, from past experience, to
know in advance which the best choice is e.g. that where the availability of
food is higher. But if all the individuals are able to do this, then all of
them will choose the same option and the availability of food will decrease
there; in that case the \textit{a priori }best option is not necessarily the
winning option. Those individuals that are not able to determine what the 
\textit{a priori} best option is will probably behave \textit{randomly} or 
\textit{persistently} (always choosing the same option). The role of
evolution and natural selection is thus expected to be crucial in these
processes, as stated in (Heesch and Little, 2006).

We note that these decision-making mechanisms are also common in human
behaviour. For instance, drivers who have to choose between two alternative
routes in order to avoid traffic jams do not analyze every past experience
and make a decision according to a pool of strategies (contrary to what is
suggested by some authors (Hod, 2003)), but mainly use simpler strategies
like \textit{persistent} behaviour (they always choose the same route
because they do not like to take risks) or they may simply listen to the
traffic news to find out what the \textit{a priori} best route is.

According to these arguments, some essential elements which are absent in
the EMG must be considered in order to get a realistic implementation of
minority games in ecology. So, the aim of this work is to present a new game
where competition for resources is also introduced by means of a minority
rule, but the dynamics and strategies followed by the agents aim to capture
the dynamics of some ecological systems. In what follows, we will refer to
this new model as the Evolutionary Learning Game (ELG).

\section{Mistiming in predator-prey systems caused by climate change}

We now introduce a specific problem that has attracted the interest of
ecologists in recent years (van Noordwijk et. al., 1995; Visser et. al.,
1998; Grieco et. al., 2002; Visser et. al., 2004; Gienapp and Visser, 2006)
and has strongly motivated our approach. In many species of bird,
individuals must face the problem of choosing the correct time for egg
laying. This choice becomes dramatic if the availability of food is
restricted to a very short period of time. So, for survival in breeding, the
correct timing of egg laying is necessary, so that the feeding period
matches the food peak. This process has been studied in recent decades for
some species, such as great tits (\textit{Parus major}) and blue tits (%
\textit{Parus caeruleus}), whose main prey (caterpillar) is only available
for two or three weeks in the late spring (Visser et. al., 2004). At the
moment of egg laying (approximately one month before), the birds do not know
when the food peak will happen. The problem is partially overcome by the way
many of these birds develop with age the ability to follow some cues (based
on climate and other environmental parameters) to predict the right time for
laying (van Noordwijk et. al., 1995; Grieco et. al., 2002; Gienapp and
Visser, 2006). In general, this capacity of an individual to adapt its
behaviour to the environmental conditions is known as phenotypic plasticity,
and is usually a heritable trait. Specifically, it has been demonstrated
(Nussey et. al., 2005) that plasticity in egg laying for birds is heritable.

The effects of global climate change, however, have put many biological
species to the test (Parmesan, 2006). As a consequence of warmer springs,
caterpillars have advanced their hatching date in many habitats (Visser et.
al., 1998; Visser et. al., 2004), so those birds with a higher plasticity in
laying are expected to adapt better to the new situation. According to the
observational data, some bird populations have become adapted, but in some
other cases a very weak response to the new situation has been observed
(Visser et. al., 2004; Gienapp and Visser, 2006). In the latter case, the
mismatching between the feeding period and the food peak will probably lead
to a decline in the number of individuals (Both et. al., 2006) or the
habitat fitness (Visser, 2007). Although different explanations have been
provided, there is no clear understanding of why different populations show
different responses to the changing conditions (Gienapp and Visser, 2006).
As we discuss below, our model provides some arguments that support the idea
that resource constraints from each specific habitat may be responsible for
these differences.

\section{Rules of the Evolutionary Learning Game}

We need to introduce two basic ideas that are missing from the original
formulation of the EMG, in order to reach a more realistic description of
ecological systems:

\textit{i)} First, reproduction and death processes must play a fundamental
role in the dynamics of the system. In the EMG the agents continue to play
indefinitely, but in ecology the consequences of choosing a wrong option can
obviously be dramatic. If we want to explore the dynamics of systems over
representative time scales, it is necessary to assume that the individuals
may disappear (die) and/or be replaced by new individuals (newborns) with
some probability. Moreover, the outcome obtained from any decision taken by
the agents must affect in some way their reproduction/survival probabilities.

\textit{ii)} Secondly, in the EMG all of the agents have access to the same
information, and so all of them may use the same strategy. In ecology,
however, as long as an individual grows up it gains experience and, in
consequence, it is expected to choose better options. So, an individual
learning process must be considered somehow. In fact, the situation where
the strategies chosen by the agents are based on their individual histories
has already been studied for the EMG (see (de Cara, 2000) and the references
therein), but here we will explore the concept of learning from a different
perspective.

In our model, each of the $N$ agents competing in the game must repeatedly
choose between options A or B. Whether or not the decision taken by the
agent is the good one will be determined by a minority rule with an
arbitrary cutoff, as defined in (Johnson et. al., 1999b). This means that we
assign a resource capacity $L$ to one of the two options (we consider $%
0<L<N/2$ without loss of generality) and a resource capacity $N-L$ to the
other option. If the real number of agents choosing the first option is
below $L$, then the resources \textit{per capita} in that option are higher
than in the other one, so those agents are the winners and the agents
choosing the other option are the losers. If the number of agents choosing
the first option is above $L$, then the contrary arguments hold.

We will consider that the option with capacity $L$ is not always the same,
but is chosen randomly every time step in order to incorporate the effects
of a fluctuating environment, so sometimes option A will be the \textit{a
priori} best option (that with a higher capacity) and sometimes not.

At each time step, the winners are rewarded with the possibility of
reproductive success. Every winner is given the possibility of producing a 
\textit{newborn} agent with a probability $r$. The newborn will replace one
of the agents in the game (to keep $N$ constant) chosen randomly, so we
assume that all of the agents are equally likely to die.

The agents choose option A or B according to the following rules. Younger
agents act \textit{persistently}: they make their first choice randomly and,
after that, they continue to choose the same option. However, after each
time step all the \textit{persistent} agents are given the possibility of 
\textit{learning} with probability $pp_{k}$ (here $pp$ stands for phenotypic
plasticity and the subindex $k$ denotes the $k$th agent). If they \textit{%
learn}, it means that they give up \textit{persistent} behaviour; from then
on, they always choose the option with a higher capacity, so we will say
that they become \textit{wise} agents.

Phenotypic plasticity in our model is thus considered equivalent to a
learning capacity. This capacity can be inherited as follows: when a \textit{%
newborn} appears, its characteristic probability $pp_{k}$ is chosen randomly
from an interval of width $w$ centred on the value of $pp_{k^{\prime }}$
from its \textit{father}, with reflecting boundary conditions at $pp_{k}=0$
and $pp_{k}=1$. So, we introduce an evolutionary dynamics for the
probabilities $pp_{k}$ into the model in a similar way as in the original
formulation of the EMG (Johnson et. al., 1999a). But note that in this case
the meaning of the width $w$ is extremely important for the dynamics of the
system, as it measures the heritability of $pp_{k}$, so the best strategies $%
pp_{k}$ will be transmitted to the \textit{newborns} only if $w$ is not too
high.

These are all the rules for our ELG. All of the agents will become \textit{%
wise} sooner or later unless they die first, but if there are too many 
\textit{wise} agents then the \textit{a priori} best option will be crowded
and will probably be the wrong one. A complex dynamic thus emerges where
learning as fast as possible is not necessarily the best strategy, which may
seem counterintuitive at first.

As some of the rules presented could be considered too simple or unrealistic
from a biological point of view, we tried to implement many different models
with increasingly complex rules in order to compare their performances:

\textit{i)} We tried to introduce explicit reproduction and death algorithms
in many different ways (for example, by using exponential or logistic
growth), so that the number of agents $N$ was allowed to change with time.

\textit{ii)} We tried to replace the minority rule with some other
competition rules, even rules that allowed all the agents to be winners (or
losers) at the same time. For instance, we considered two independent
capacities $L_{A}$ and $L_{B}$ for the two possible options, so if the
number of individuals choosing option A is above (below) $L_{A}$ those
agents are considered losers (winners).

\textit{iii)} We tried to reward and/or fine agents on their reproductive
success and/or their probability of survival. Of course, it is not necessary
for the reproductive success to be completely suppressed for the losers as
in the simplified version we have described; we could consider two
reproductive rates $r_{w}$ and $r_{l}$ for winners and losers respectively,
with $r_{w}>r_{l}$.

\textit{iv)} We tried to consider that the switching from \textit{persistent}
to \textit{wise} behaviour is not so radical, but the agents learn
progressively according to a rate given by $pp_{k}$.

After these and many other trials, we have found that the qualitative
behaviour exhibited by the ELG (which is shown in the following Section) is
highly robust. According to our results, it seems that there are only two
elements which are strictly necessary in order to obtain that behaviour: 
\textit{i)} a learning process regulated by the probabilities $pp_{k}$ and 
\textit{ii)} that the number of agents rewarded (fined) is proportional to
the number of winners (losers). The version of the ELG we have presented
here is one of the simplest possible, and so it offers the advantage that
some analytical treatment is possible, as we will show below.

\section{Results}

The greatest interest of our model lies in the form of the distribution of
phenotypic plasticities $P(pp_{k})$ that is reached in the steady state. In
Figure 1 we summarize the behaviour of $P(pp_{k})$ as a function of the
three parameters of the model: $L$, $r$ and $w$. All the results shown here
were obtained by computing the form of $P(pp_{k})$ for $N=2001$ after 10000
time steps (which is far enough to reach the steady state), and carried out
an average of 25 different realizations. Initially all the agents were
considered \textit{newborns} and the values of $pp_{k}$ were assigned
randomly; anyway, we have checked that our results are independent of the
initial conditions chosen.

The series of plots from \textit{1.a} to \textit{1.d} shows how $P(pp_{k})$
changes when the value of $L$ is modified. Figures \textit{1.a} and \textit{%
1.d} correspond to extreme situations that are clearly predictable. In the
first case, when $L\rightarrow N/2$ both options A and B have similar
resource capacities. Therefore, performing as a \textit{wise} agent does not
represent an advantage, because the resource of choice (i.e. the larger one)
reaches its smallest possible size leading to overcrowding among wise
agents. As a consequence, learning is avoided and there is a tendency $%
pp_{k}\rightarrow 0$. In the regime $L\rightarrow 0$ one of the two options
is much better than the other one. In this situation, performing as a 
\textit{wise} agent is a strong advantage, and so the tendency $%
pp_{k}\rightarrow 1$ should be expected. But, surprisingly, there is a wide
range of intermediate values of $L$ where segregated (obviously asymmetric)
behaviour is found. This means that in intermediate situations the dynamics
of the system tends to favour individuals which either learn as fast as
possible or avoid learning as much as possible.

Although segregated behaviour was also found for the EMG, the situation
reported here is clearly different. In the case of the EMG\ (Johnson et.
al., 1999a) the segregated behaviour in the steady state was independent of
the initial conditions and the values of the parameters introduced. However,
Hod and Nakar (2002) proved later that the model is extremely sensitive to
the prize-to-fine ratio, so for some parameters one observes a sharp
transition where self-segregation is destroyed. On the contrary, we have not
noticed such effects in our model, but the form of $P(ppk)$ always changes
smoothly for any region of parameters considered. The other main difference
between the EMG and the ELG is that the results obtained here show an
asymmetric distribution of $P(pp_{k})$. The reason for this is that in the
ELG the \textit{persistent} and \textit{wise} agents do not necessarily
choose different options (while in the EMG $p_{k}\rightarrow 0$ and $%
p_{k}\rightarrow 1$ represent opposite behaviours). This, together with the
different backgrounds considered and some ideas discussed below, shows that
the general dynamics of the EMG and the ELG are different, although there
are some major similarities between both.

One can observe in the series \textit{1.e} to \textit{1.h} the role of the
reproduction probability $r$ on $P(pp_{k})$ while keeping the other
parameters constant. A high value of $r$ involves the appearance of many 
\textit{newborns} and, according to the discussion above, a \textit{wise}
strategy will then perform better that in a situation with few \textit{%
newborns}. For this reason, the model shows a tendency $pp_{k}\rightarrow 0$
for low $r$ and a tendency $pp_{k}\rightarrow 1$ for high $r$. In
intermediate situations, a segregated distribution is found again.

Finally, the role of $w$ is shown in plots from \textit{1.i} to \textit{1.l}%
. As discussed above, the value of $w$ determines the heritability of the
phenotypic plasticity. Low values of $w$ represent a high level of
heritability and so the best strategies persist, while a high value of $w$
means that best strategies are not well transmitted to breeding; so, for the
latter $P(pp_{k})$ is expected to tend to be uniform. The value of $w$ also
indirectly affects the number of winners and losers; if $w$ is too high the
best strategies do not persist and then the average number of winners
decreases. For this reason it is difficult to predict the exact form of $%
P(pp_{k})$ as a function of $w$. Actually, the specific role of $w$ in the
game is fairly complicated, so this point will be addressed in detail in a
further study. Note that this is another important difference from the case
of the EMG, where the final distribution $P(p_{k})$ is almost independent of 
$w$ (Johnson et. al., 1999a).

We can give some analytical support to the results shown in Figure 1 by
means of a mean-field-like approach as proposed before for the EMG (Lo 
\textit{et. al.}, 2000). Here we keep as much as possible to the notation
used there in order to facilitate understanding.

First of all, note that we know the option that the \textit{wise} agents
will choose. We also know that every \textit{persistent} agent made its
first choice randomly, so we should expect on average for half of them to
choose A and the other half to choose B. Therefore, the whole problem is
reduced to finding out how many \textit{persistents} are in the game. We
denote $F_{N}(n)$ as the probability of $n$ of the $N$ agents in the game
being \textit{persistent}. Similarly, we define $G_{N-1}^{k}(n)$ as the
probability of $n$ of the agents being persistent, given that the $k$th
agent is the only one that has not made its choice yet. Then, the following
relation holds: 
\begin{equation}
F_{N}(n)=\Gamma _{pp_{k}}G_{N-1}^{k}(n-1)+\left( 1-\Gamma _{pp_{k}}\right)
G_{N-1}^{k}(n),  \label{1}
\end{equation}
where $\Gamma _{pp_{k}}$ is the probability of the $k$th agent being \textit{%
persistent}, given that its phenotypic plasticity is $pp_{k}$. In the
following, we will use the simplified notation $\Gamma \equiv \Gamma
_{pp_{k}}$.

On the other hand, the winning probability $\tau _{pp_{k}}$ of an agent that
has a plasticity $pp_{k}$ can be written as 
\begin{equation}
\tau _{pp_{k}}=\frac{\Gamma }{2}\sum_{n=0}^{\alpha -1}G_{N-1}^{k}(n)+\left(
1-\frac{\Gamma }{2}\right) \sum_{n=\alpha +1}^{N-1}G_{N-1}^{k}(n),  \label{2}
\end{equation}
where $\alpha =Int(2L)$ denotes the integer part of $2L.$ Following now the
same treatment as in (Lo et. al., 2000), we get from (\ref{1}) and (\ref{2})
the expression 
\begin{eqnarray}
\tau _{pp_{k}} &=&\frac{\Gamma }{2}\sum_{n=0}^{\alpha }F_{N}(n)+\left( 1-%
\frac{\Gamma }{2}\right) \sum_{n=\alpha +1}^{N-1}F_{N}(n)  \nonumber \\
&&+\Gamma (\Gamma -3/2)G_{N-1}^{k}(\alpha ).  \label{3}
\end{eqnarray}

As stated in (Lo et. al., 2000), the two first terms correspond to the
winning probability of a $k$th agent whose action does not modify the result
of the game, once the other $N-1$ agents have made their choice. Hence, the
third term is the essential one, as it measures the influence of the $k$th
agent on the final result. For example, imagine that after the first $N-1$
agents have made their choice, one half of them decide to follow option A
(so one half choose B); then, the last agent?s decision will determine which
the winning option is; the influence of this final decision on the
probability $\tau _{pp_{k}}$ is what the last term in (\ref{3}) measures. In
order to analyze this term, we need to find the explicit expression of $%
\Gamma $ as a function of $pp_{k}$. This is easy to do, because after each
time step the \textit{persistents} have a probability $pp_{k}$ of becoming 
\textit{wise} agents and an average survival probability $s$ (according to
the rules of our ELG, only the range $0.5<s<1$ holds here). Then, the
probability of the $k$th agent being \textit{persistent} is 
\begin{equation}
\Gamma _{pp_{k}}=\frac{\sum_{i=0}^{\infty }\left[ s^{i}(1-pp_{k})^{i}\right] 
}{\sum_{i=0}^{\infty }s^{i}}=\frac{1-s}{1-s(1-pp_{k})}.
\end{equation}

Now, we can give an explicit expression for the term $\Gamma (\Gamma -3/2)$
from Equation (\ref{3}). It can be seen that $\Gamma (\Gamma -3/2)$ has the
same appearance throughout the whole range $0<pp_{k}<1$ and for the proper
range of survival probabilities $0.5<s<1$. It is always a negative convex
function with a relative minimum at $pp_{k}=(1-s)/3s$. This means that the
third term in (\ref{3}) always tends to favour extreme values of $pp_{k}$,
which facilitates the emergence of segregated behaviour; so, this gives some
justification to our numerical results.

We stress, however, that stationary analytical approaches such as the one
used here have some limitations, as the EMG and similar models never reach a
true stationary distribution (Hod, 2002). In fact, we have found for the ELG
that the number of \textit{persistents} oscillates periodically over time,
in accordance with similar results found for the EMG (Hod, 2002). In that
work, it was also argued that when the amplitude of these oscillations
increases, we observe in the EMG a transition from self-segregated behaviour
to clustering (where clustering is characterized by a single-peaked
distribution of $p_{k}$ values around $p_{k}=1/2$). It is interesting to
note that, on the contrary, in the ELG the amplitude of the oscillations
increases in the region where self-segregated behaviour is found; so, the
oscillations in our model seem to enforce self-segregation rather than
destroying it. This situation is shown in Figure 2, where the number of 
\textit{persistents} in the steady regime is plotted as a function of time
for different situations; dotted, dash-dotted and solid lines correspond to
the cases \textit{1.a}, \textit{1.c} and \textit{1.d} reported in Figure 1,
respectively. From Figure 2, it is also clear that the mean number of 
\textit{persistents} in the game increases as $L$ decreases, in accordance
with our discussion above.

\section{Discussion}

We have presented a model that sets the ideas of minority games, which are
considerably popular tools for describing competition for resources in
economics, into an ecological context. The model presented here shows how
social segregation emerges from an evolutionary learning process (determined
by the distribution $P(pp_{k})$) in a group of individuals competing for
strongly limited resources. Note that if the learning process were not
introduced to the model, then the dynamics of the system would be trivial.
Neither persistent nor wise behaviour on its own is an efficient strategy;
evolutionary learning is the key ingredient here for finding efficient
cooperation between both. This idea, together with the robustness shown by
our model (many other implementations with more realistic rules led to
similar qualitative results) seems to suggest that our model could be of
interest for understanding social organization in complex evolutionary
systems.

The results obtained here show that in the situation described by our model
intermediate learning strategies cannot persist for long times; the tendency
is always towards improvement ($pp_{k}\rightarrow 1$), suppression ($%
pp_{k}\rightarrow 0$) or the coexistence of both (segregated distribution).
It also contradicts the intuitive idea that those individuals with a higher
learning capacity must always be favoured by selection. This is a
consequence of the strong competition process which is assumed in our ELG
and in minority games in general: sometimes the \textit{a priori} worst
option can be the best because many agents tend to choose the \textit{a
priori} best option and then competition for the latter is higher.

Finally, we come back to the problem of egg laying in birds described
before, which can now be addressed using the ideas about learning and
phenotypic plasticity discussed here. In order to provide an analogy with
our model, we could imagine that option A means laying early and option B
means laying later. Those individuals with a higher plasticity very quickly
become able to follow some environmental cues in order to predict the right
option A or B. However, if the penalty for choosing a wrong option is not
high (because there are some other food resources available, or there are
some other environmental constraints on laying...) then selection will not
favour individuals with higher plasticity. In that case, when the
individuals must face sudden environmental changes their capacity to respond
will be weak. In those habitats where phenotypic plasticity is strongly
rewarded (i.e., for $L$ small in our model), individuals are expected to
follow efficient learning strategies, so they will be able to respond better
to environmental changes.

Therefore, the results of our model provide an evolutionary basis to the
idea that environmental constraints from each specific habitat could explain
why different bird populations are responding differently to climate-driven
changes in the behaviour of their prey, as recent experiments have suggested
(Gienapp and Visser, 2006). However, empirical evidence supporting our ideas
about learning and phenotypic plasticity is still lacking. We believe that
it would be of major interest if experimentalists were to attempt to check
the predictions made by our model in real ecological systems.

\section*{Acknowledgements}

This research has been partially supported by the Generalitat de Catalunya
through grant 2006-BP-A-10060 (DC), by the project CGL 2007-60797 (JELl) and
by grants FIS 2006-12296-C02-01, SGR 2005-00087 (VM).

\section*{References}

Both, C., Bouwhuis, S., Lessells, C.M., Visser, M.E., 2006. Climate change
and population declines in a long-distance migratory bird. Nature 441, 81-83.

Challet, D., Zhang, Y.C., 1998. Emergence of cooperation and organization in
an evolutionary game. Physica A 246, 407-418; On the minority game:
Analytical and numerical studies. Physica A 256, 514-532.

Cont, R., Bouchaud, J.P., 2000. Herd behavior and aggregate fluctuations in
finanical markets. Macroeconom. Dyn. 4, 170-196.

de Cara, M.A.R., Pla, O., Guinea, F., 2000. Learning, competition and
cooperation in simple games. Eur. Phys. J. B. 13, 413-416.

Gienapp, P., Visser, M.E., 2006. Possible fitness consequences of
experimentally advanced laying dates in Great Tits: differences between
populations in different habitats. Funct. Ecol. 20, 180-185.

Grieco, F., van Noordwijk, A.J., Visser, M.E., 2002. Evidence for the effect
of learning on timing of reproduction in blue tits. Science 296, 136-138.

Heesch, D., Little, M., 2006. Decision-making in variable environments - a
case of group selection and inter-generational conflict? Theor. Pop. Biol.
69, 121-128.

Hod, S., 2003. Time-dependent random walks and the theory of complex
adaptive systems. Phys. Rev. Lett. 90, 128701-4.

Hod, S., Nakar, E., 2002. Self-segregation versus clustering in the
evolutionary minority game. Phys. Rev. Lett. 88, 238702-4.

Hod, S., Nakar, E., 2004. Evolutionary minority game: The roles of response
time and mutation threshold. Phys. Rev. E 69, 066122.

Johnson, N.F., Hui, P.M., Jonson, R., Lo., T.S., 1999. Self-Organized
segregation within an evolving population. Phys. Rev. Lett. 82, 3360-3363.

Johnson, N.F., Hui, P.M., Zheng, D., Thai, C.W., 1999. Minority game with
arbitrary cutoffs. Physica A 269, 493-502.

Johnson, N.F., Leonard, D.J.T., Hui, P.M., Lo., T.S., 2000. Evolutionary
freezing in a competitive population. Physica A 283, 568-574.

Johnson, N.F., Jefferies, P., Hui, P.M., 2003. Financial market complexity.
Oxford University Press, Oxford.

Lo, T.S., Hui, P.M., Johnson, N.F., 2000. Theory of the evolutionary
minority game. Phys. Rev. E 62, 4393-4396.

Nowak, M.A., Sasaki A., Taylor, C., Fudenberg, D., 2004. Emergence of
cooperation and evolutionary stability in finite populations. Nature 428,
646-650.

Nowak, M.A., 2006. Five Rules for the Evolution of Cooperation. Science 314,
1560-1563.

Nussey, D.H., Postma, E., Gienapp, P., Visser, M.E., 2005. Selection on
heritable phenotypic plasticity in a wild bird population. Science 310,
304-306.

Parmesan, C., 2006. Ecological and evolutionary responses to recent climate
change. Annu. Rev. Ecol. Evol. Syst. 37, 637-669.

Sysi-Aho, M., Chakraborti, A., Kaski, K., 2003. Biology helps you to win a
game. Phys. Scr. T106, 32-35.

Tella, J.A., de Cara, M.A.R., Pla, O., Guinea F., 2001. A model for
predation pressure in colonial birds. In Sixth Granada Lectures in
Computational Systems. Ed. Garrido, P.L., Marro, J. AIP Conf. Proc. 574.
AIP, New York.

van Noordwijk, A.J., McCleery, R.,Perrins, C., 1995. Selection for the
timing of great tit breeding in relation to caterpillar growth and
temperature J. Anim. Ecol. 64, 451-458.

Visser, M.E., van Noordwijk, A.J., Tinbergen, J.M., Lessells, C.M., 1998.
Warmer springs lead to mistimed reproduction in great tits (Parus major).
Proc. Royal Soc. London B 265, 1867-1870.

Visser, M.E., Both, C., Lambrechts, M.M., 2004. Global climate change leads
to mistimed avian reproduction. Adv. Ecol. Res. 35, 89-110.

Visser, M.E., 2007. Keeping up with a warming world; assessing the rate of
adaptation to climate change. Proc. R. Soc. B. DOI 10.1098/rspb.2007.0997.

Quan, H.-J., Wang, B.-H., Hui, P.M., Luo, X.-S., 2003. Self-segregation and
enhanced cooperation in an evolving population through local information
transmission. Phys. A 321, 300-308.

\newpage

\textbf{Figure captions}

Figure 1.

Plot of the distribution $P(pp_{k})$ in the steady state (after 10000 time
steps) for $N=2001$ averaged over 25 realizations. The series \textit{a} to 
\textit{d} shows the effect of varying $L$ while keeping $w=0.1$ and $r=0.4$%
. The series \textit{e} to \textit{h} shows the effect of varying $r$ while
keeping $L=0.2N$ and $w=0.1$. The series \textit{i} to \textit{l} shows the
effect of varying $w$ while keeping $L=0.2N$ and $r=0.3$.

Figure 2.

Number of \textit{persistents} as a function of time once in the steady
state (the results plotted correspond to times $500<t<1000$ and $N=2001$).
The dotted line corresponds to the parameter values in Figure \textit{1.a.};
the dashed-dotted line corresponds to parameter values in Figure \textit{1.c.%
}; the solid line corresponds to the parameter values in Figure \textit{1.d.}

\end{document}